\documentclass[aps,prl,twocolumn,superscriptaddress]{revtex4}

\usepackage{graphicx}
\usepackage{bm}
\usepackage{amsmath}
\usepackage{amssymb}
\usepackage{hyperref}
\usepackage{natbib}

\begin{document}

\title{Compression of nanowires using a flat indenter: Elasticity measurement in nanoscale}

\author{Zhao Wang}
\email{wzzhao@yahoo.fr}
\affiliation{Empa, Swiss Federal Laboratories for Materials Testing and Research, Laboratory for Mechanics of Materials and Nanostructures, Feuerwerkerstrasse 39, CH-3602 Thun, Switzerland.}
\affiliation{LITEN, CEA-Grenoble, 17 rue des Martyrs, 38054 Grenoble Cedex 9, France.}

\author{W.M. Mook}
\affiliation{Empa, Swiss Federal Laboratories for Materials Testing and Research, Laboratory for Mechanics of Materials and Nanostructures, Feuerwerkerstrasse 39, CH-3602 Thun, Switzerland.}

\author{Ch. Niederberger}
\affiliation{Empa, Swiss Federal Laboratories for Materials Testing and Research, Laboratory for Mechanics of Materials and Nanostructures, Feuerwerkerstrasse 39, CH-3602 Thun, Switzerland.}

\author{R. Ghisleni}
\affiliation{Empa, Swiss Federal Laboratories for Materials Testing and Research, Laboratory for Mechanics of Materials and Nanostructures, Feuerwerkerstrasse 39, CH-3602 Thun, Switzerland.}

\author{L. Philippe}
\affiliation{Empa, Swiss Federal Laboratories for Materials Testing and Research, Laboratory for Mechanics of Materials and Nanostructures, Feuerwerkerstrasse 39, CH-3602 Thun, Switzerland.}

\author{J. Michler}
\affiliation{Empa, Swiss Federal Laboratories for Materials Testing and Research, Laboratory for Mechanics of Materials and Nanostructures, Feuerwerkerstrasse 39, CH-3602 Thun, Switzerland.}

\begin{abstract}
A new experimental approach for the characterization of the lateral elastic modulus of individual nanowires is demonstrated by implementing a micro/nano scale diametrical compression test geometry, using a flat punch indenter inside of a scanning electron microscope (SEM). A 250 nm diameter single crystal silicon nanowire is tested. Since silicon is highly anisotropic, the compression axis of the wire was determined by electron backscatter diffraction (EBSD). A two dimensional analytical closed-form solution based on a Hertz model is presented. The results of the analytical model are compared with those of finite-element simulations and to the experimental diametral compression results and show good agreement. (revised manuscript)
\end{abstract}

\maketitle
Nanowires have attracted great interest in the past decade due to their interesting mechanical, electrical, and optical properties \cite{Wu2005,Cui2003}. This has enabled nanowires to gain advantages in a wide range of applications such as nanoelectromechanical systems \cite{Craighead2000}, gas sensors \cite{Wan2004}, resonators \cite{Kong2003a}, light sensors \cite{Huang2005}, photovoltaic cells \cite{Jamil2009}, and atomic force microscopy and scanning tunneling microscopy tips \cite{Becker2008}. In this paper we focus on the characterization of the mechanical properties of nanowires, in particular by in-situ scanning electron microscopy (SEM) diametral compression using a flat punch. 

The main experimental techniques employed to determine the mechanical properties of nanowires are mechanical resonant methods \cite{Philippe2007,Wang2001}, AFM-based bending methods \cite{Wu2005,Hoffmann2006,Hoffmann2007}, tensile testing \cite{Zhang2009}, and nanoindentation methods \cite{Feng2006,Li2003,Mao2003}. As pointed out by several research groups, nanowire indentation has the great advantage of little specimen preparation, in fact the nanowire does not need to be held in place as with all the other techniques. Not only does the holding require intensive preparation but it can also influence the measured mechanical response since the mechanical and adhesive properties of the holding element are generally neglected. Nanoindentation has a couple of drawbacks such as the alignment of the nanoindenter tip with respect to the nanowire \cite{Ghisleni2009} and the non-applicability of the Oliver and Pharr method \cite{Chang2009,Shu2009} since the nanowire cannot be considered a half-space\cite{Oliver1992}.

To overcome these issues, it is introduced here the in-situ scanning electron microscope (SEM) diametral compression test performed at the mirco/nano scale. In this approach, the misalignment defined as the non-perfect perpendicularity between the nanowire axis and the normal to the flat punch is attenuated since the experiments are conducted in a SEM while the misalignment due to the non-intersection of the nanowire axis and the flat punch axis is not as critical as in the case of wire indentation \cite{Ghisleni2009} since the indenter used is a flat punch with a diameter that is much larger than the diameter of the nanowire. Concerning the diametrical compression data analysis, Hertz theory is well suited. This allows one to solve, at least in the 2D simplified case described in the next section, in closed-form this problem allowing one to determine the lateral elastic modulus of the wire from the load-displacement data obtained experimentally. In order to validate this technique, the experimental test was compared to an analytically closed-form and a numerical solution of the diametrical compression test problem. 

To the authors knowledge this is the first time that a wire is compressed by a flat punch in the micro/nano scale. This test geometry is analogus to the macro-scale Brazilian test \cite{Hondros1959,Cheng1998} that was introduced back in 1950s to study ceramic fracture behavior.

The remainder of this paper is organized as follows. In Sec.II we first describe the formulation of our analytical approach. The results of the analytical model, experiments and numerical simulations are compared and discussed in Sec.III, followed by conclusions in Sec.IV.

In this section, we outline a closed-form Hertzian solution to determine the elastic modulus of the nanowire tested with a flat indenter, analyzing the load-displacement ($P-\delta$) relationship. We study the cross section of the system in two physical dimensions. The total displacement $\delta$ measured from the compression experiment consists of three terms: $\delta=\delta_{i}+\delta_{w}+\delta_{s}$ where the subscripts $i$, $w$ and $s$ indicate the indenter, the nanowire and the substrate, respectively. 

\begin{figure}[ht]
\centerline{\includegraphics[width=9cm]{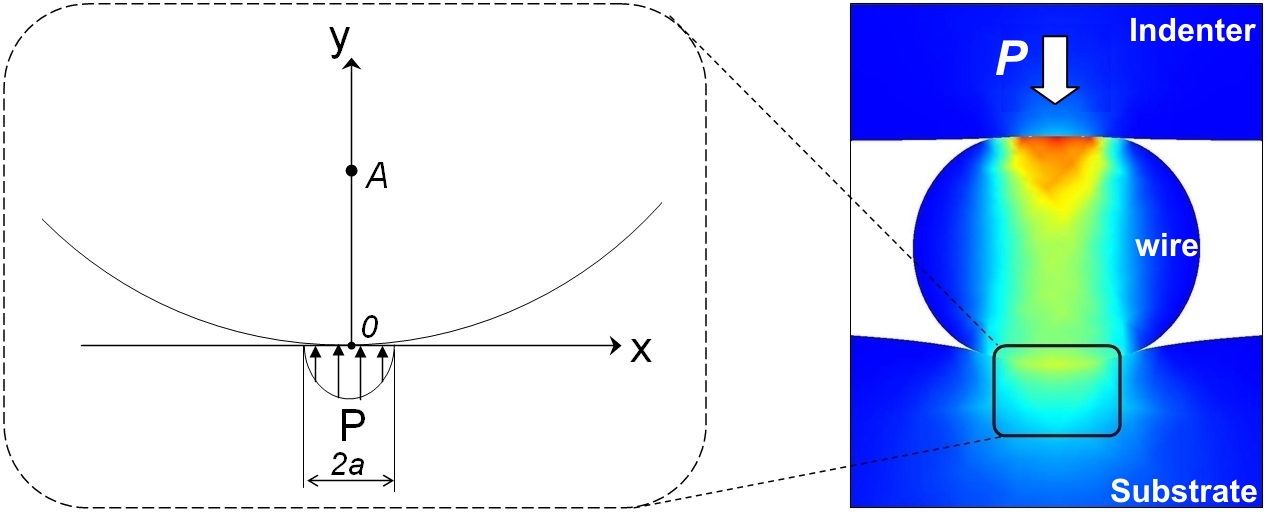}}
\caption{\label{fig:schema}
2D diagram of an indentation test on a nanowire (cylindrical cross section).}
\end{figure}

To calculate the deformation, we seek the two components of stress ($\sigma_{x},\sigma_{y}$) distributed at an arbitrary point $A(x,y)$ at $x=0$ in the cylinder (of radius $R$) under a pressure $P$ (Figure\,\ref{fig:schema}). By assuming a Hertzian traction distribution ($=2P\sqrt{1-x^{2}/a^{2}}/a\pi$ and $R>>a$) at the interface \cite{Johnsonbook}, the stress can be written as

\begin{equation}
\label{eq:1}
\begin{array}{c}
\left\{ \begin{array}{c}
\sigma_{x}= \frac{P}{\pi} \left[ \frac{1}{R}-\frac{2(a^{2}+2y^{2})}{a^{2}\sqrt{a^{2}+y^{2}}} + \frac{4y}{a^{2}} \right] \\
\sigma_{y}= \frac{P}{\pi} \left[ \frac{1}{R}-\frac{2}{2R-y} + \frac{2}{\sqrt{a^{2}+y^{2}}} \right]
\end{array}\right.\\
\end{array}\,
\end{equation}

\noindent where $a$ is the transverse semi-contact length (Figure\,\ref{fig:schema}), which can be calculated according to a Hertzian distribution of pressure at the interface as

\begin{equation}
\label{eq:11}
a=\sqrt{4PR/\pi E^{int}}
\end{equation}

\noindent where $E^{int}$ stands for the interface elastic modulus, which is defined as 

\begin{equation}
\label{eq:7}
\begin{array}{c}
\left\{ \begin{array}{c}
1/E_{iw}^{int}= 1/{E_{i}^{*}}+{1}/{E_{w}^{*}}  \\
{1}/{E_{sw}^{int}}= {1}/{E_{s}^{*}}+{1}/{E_{w}^{*}} 
\end{array}\right.\\
\end{array}\,.
\end{equation}

\noindent where the subscripts $iw$ and $sw$ indicate the indenter-wire and the wire-substrate interfaces, respectively. $E^{*}=E/(1-v^{2})$ where $E$ is the elastic modulus and $v$ is the Poisson's ratio.

The vertical deformation $\epsilon_{y}$ at $A(x=0,y)$ under biaxial stress is 

\begin{equation}
\label{eq:2}
\epsilon_{y} = \frac{1}{E^{*}} (\sigma_{y}- \sigma_{x} \frac{v}{1-v}) \quad .
\end{equation}

The compression of the wire $\delta_{w}$ is given by integrating $\epsilon_{y}$ in each half cylinder $\int_{0}^{R} \epsilon_{y} dy$. The result of the integration (using Eqs.(1-5)) gives

\begin{equation}
\label{eq:6}
\delta_{w}  = \frac{2P}{\pi E^{*}_{w}}(\ln\frac{4R}{a_{i}}+\ln\frac{4R}{a_{s}}-1)  \quad .
\end{equation}

Additionally, the penetration of the nanowire into the indenter and the substrate refers to a typical problem of calculating the compression of a half-space under a Hertzian pressure, assuming that the dimension of the substrate and the indenter is much larger than that of the contact interface. A classical solution \cite{Johnsonbook} to this problem involves integration along the depth of the substrate ($h_{s}$) or indenter ($h_{i}$). It yields: 

\begin{equation}
\label{eq:5}
\begin{array}{c}
\left\{ \begin{array}{c}
\delta_{i}=\frac{P}{\pi E^{*}_{i}} ( 2\ln{\frac{2h_{i}}{a_{i}}} - \frac{v_{i}}{1-v_{i}}  )     \\
\delta_{s}=\frac{P}{\pi E^{*}_{s}}  ( 2\ln{\frac{2h_{s}}{a_{s}}} - \frac{v_{s}}{1-v_{s}}  )
\end{array}\right.\\
\end{array}\,
\end{equation}

\noindent where  $h$ is thickness of the indenter or the substrate.

In summary, the load-displacement relationship can be expressed as

\begin{equation}
\label{eq:8}
P = \delta \pi \left[ \frac{1}{E^{*}_{i}} (2\ln{\frac{2h_{i}}{a_{i}}} - \frac{v_{i}}{1-v_{i}}) + \frac{2}{E^{*}_{w}}(\ln\frac{4R}{a_{i}}+\ln\frac{4R}{a_{s}}-1) +  \frac{1}{E^{*}_{s}} (2\ln{\frac{2h_{s}}{a_{s}}} - \frac{v_{s}}{1-v_{s}}) \right]^{-1} .
\end{equation}

It is worth noting that this relationship between the load $P$ and the displacement $\delta$ is not totally linear since $a$ depends on $P$.

The compression of nanowires is simulated using a 2D finite-element (FE) model realized with a commercial computation code \cite{Comsol}. The geometry of the simulated system is shown in Figure\,\ref{fig:schema} \textit{right}. The bottom of the substrate is fixed in the simulations and an imposed vertical force (downward) is applied to the top of the punch indenter. The object sizes and the material properties of the simulation system are set with respect to our experimental setup in nanoscale. For the mesh, a plane-strain Lagrangian model is used (approxiamatly 40000 triangle meshes). These interface model is set to be the general contact model without friction.

The 250 nm diameter Si nanowire used in our experiments was prepared by the VLS method using chemical vapor deposition (CVD) with silane. The substrate was heated to about 580$^o$C at a pressure of $2 \times 10^{-7}$ mbar for 10 minutes. The temperature was then reduced to 510$^o$C and a mixture of Ar ($10$ sccm) and SiH$_4$ ($5$ sccm) was introduced for the nanowire growth for 15 minutes at a pressure of 2 mbar.

The nanowires were compressed with a Hysitron PicoIndenter$^{\circledR}$ that was built for SEM in-situ experiments \cite{Rzepiejewska-Malyska2008}. The experiments were conducted in a Zeiss DSM 962 SEM operated at an accelerating voltage of 10.0 kV. All compressions were conducted such that the substrate tilt was between 8$^o$ to 10$^o$ with respect to the electron beam. The tip was a boron-doped diamond that was focus ion beam (FIB) milled to create a flat-punch geometry. The nominal indenter tip displacement rate was $2$ nm/s using a closed feedback-loop which adequately represents displacement control during elastic segment of the compressions. An SEM image of the experimental configuration can be seen in Fig. 2(a). 

\begin{figure}[ht]
\centerline{\includegraphics[width=9cm]{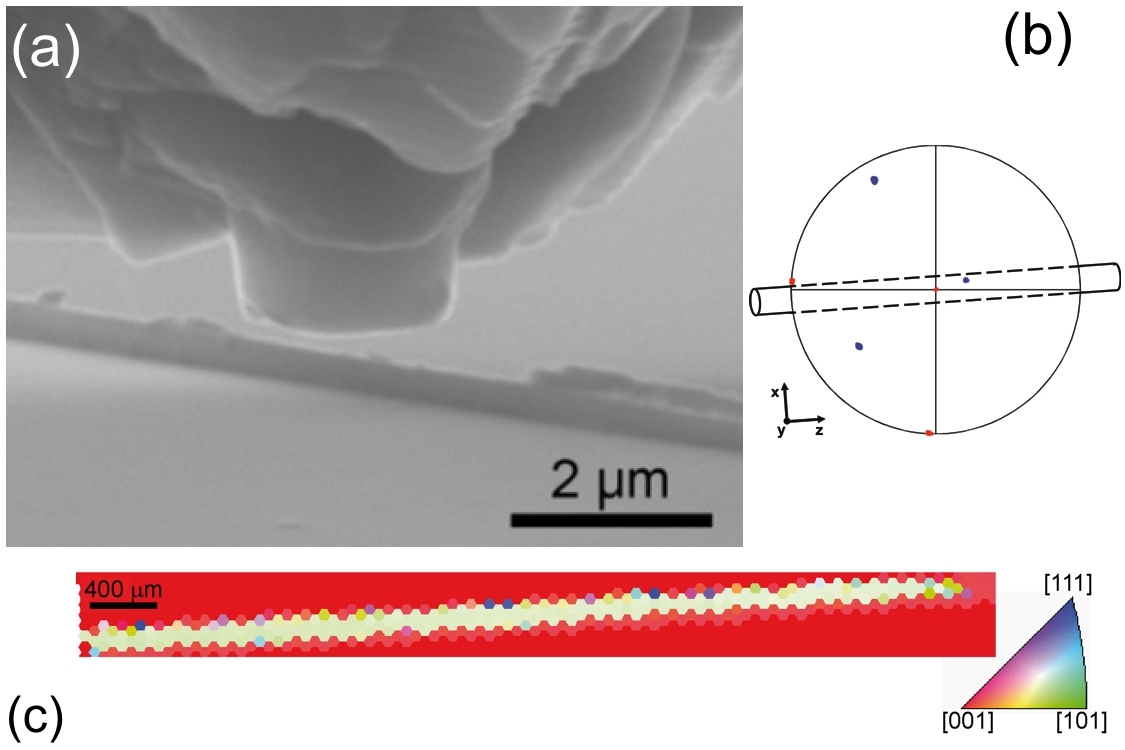}}
\caption{\label{fig:exp}
(Color online) (a)  SEM image showing a nanowire lying on the Si wafer substrate and the diamond flat-punch indenter tip. (b) Streographic projection of the $<$100$>$ directions illustrating the crystal orientation of the Si nanowire (blue dots) and the substrate (red dots). (c) EBSD mapping of the Si nanowire and the substrate. The color code gives the crystallographic orientation of the surface normal.
}
\end{figure}

The analytical model (Eq.\,\ref{eq:8}) for the deformation of the nanowire requires accurate values of the Young's Modulus $E$ and the Poisson coefficient $v$ for all involved materials. However, the elastic behavior of single crystal silicon is anisotropic and strongly depends on the orientation of the crystal \cite{Wortman1965}. In order to determine the effective values of $E$ and $v$ in our experiments, the crystallographic orientation of the compression axis of the specific nanowire has been determined by means of electron backscatter diffraction (EBSD), which can be used to identify the local crystal orientation with a spatial resolution as good as $10$ nm and an angular orientation resolution better than $0.5^{o}$ \cite{Dingley2004}. The EBSD analysis was conducted on a Hitachi S-4800 cold field emission high resolution SEM equipped with a TSL/EDAX EBSD system at an acceleration voltage of $20.0$ kV. The crystal orientation mappings of the Si nanowire and the underlying Si wafer can be seen in Figure\,\ref{fig:exp}. 

\begin{figure}[ht]
\centerline{\includegraphics[width=9cm]{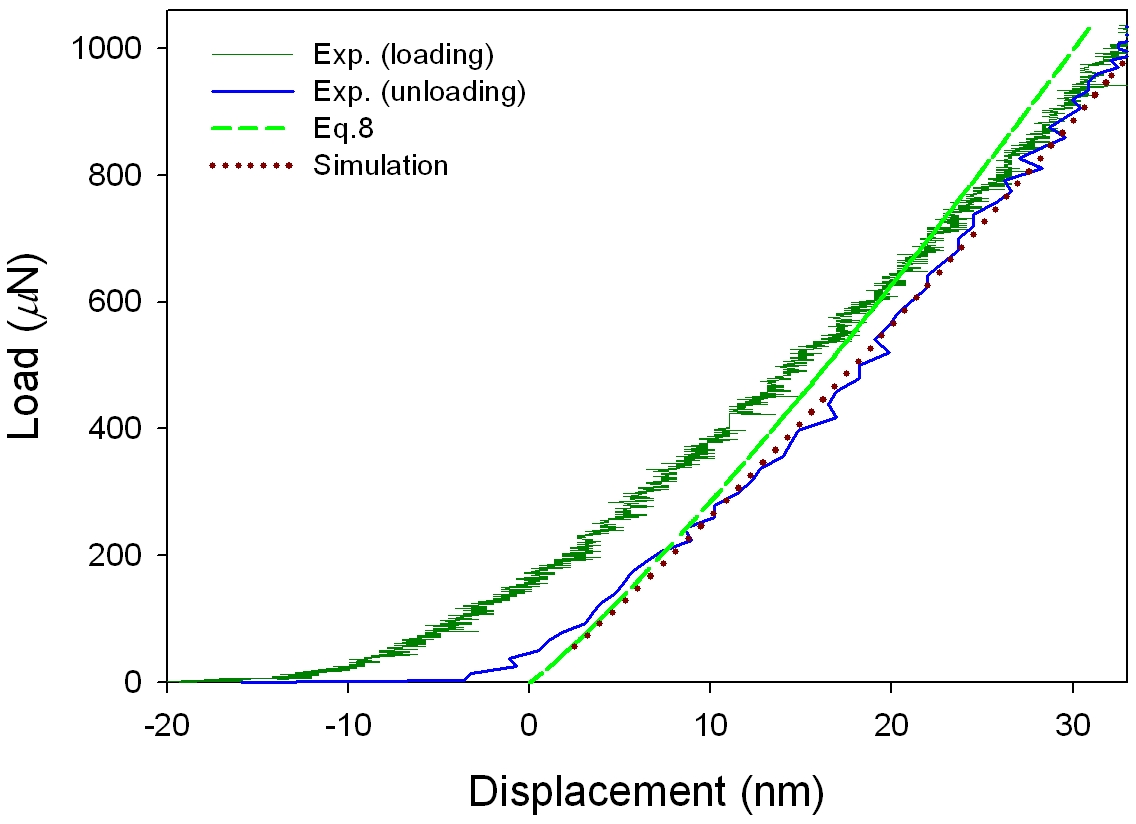}}
\caption{\label{fig:expvtheo}
(Color online) Comparison of the load-displacement curves from a nanoindentation experiment (load: Elastoplastic and unload: Elastic), the analytical model (Eq.\,\ref{eq:8}) and a FE simulation.}
\end{figure}

Two consecutive compressions were conducted on the wire. The first compression was elastic such that there was minimal difference between the loading and unloading slopes of the load-displacement graph seen in Figure\,\ref{fig:expvtheo}. A second compression was then conducted at which point plasticity occurred. The plasticity enabled the length of the contact to be estimated as $1.34$ $\mu$m by subsequent SEM plan-view imaging of the deformed region of the nanowire. This longitudinal contact length is very important since it must be used to normalize the load, $P$, in Eq.\,\ref{eq:8}. It should be noted that the contact length within the first $10$ nm of displacement in Figure\,\ref{fig:expvtheo} is not constant, but increases to the final recorded length of $1.34$ $\mu$m. This is due to small but unavoidable misalignments between the flat punch, nanowire and substrate. Excluding these first few nanometers, the remainder of the elastic portion of the first compression is then used in order to fit the effective elastic modulus of the nanowire using Eq.\,\ref{eq:8}. 

The EBSD mapping confirms that our Si nanowire is a single crystal. The crystal orientation of the nanowire and the underlying substrate are visualized in the pole figure of Figure\,\ref{fig:exp}. The $<$100$>$ crystal directions of the nanowire (red) and the Si wafer substrate (blue) are displayed in a stereographic projection in Figure\,\ref{fig:exp}(b). The configuration of the nanowire is drawn schematically into the pole figure. The crystal orientation of the nanowire is such that the compression direction y expressed in the crystal coordinate system is (0.17, 0.92, 0.34). The lateral direction x is (0.87, -0.39, 0.29). According to the measured crystal orientations, the Young's modulus and the Poisson's ratio of the nanowire are estimated to be $E_{w} = 147$ GPa and $v_{w}$=0.233 based on the data of Wortman and Evans \cite{Wortman1965}. In the substrate, the principal direction of the stress varies and strongly depends on the position within the substrate. Therefore, based on indentation results, an average value of $153$ GPa has been assumed for the modulus of the substrate \cite{Cook2006}.

\begin{figure}[ht]
\centerline{\includegraphics[width=9cm]{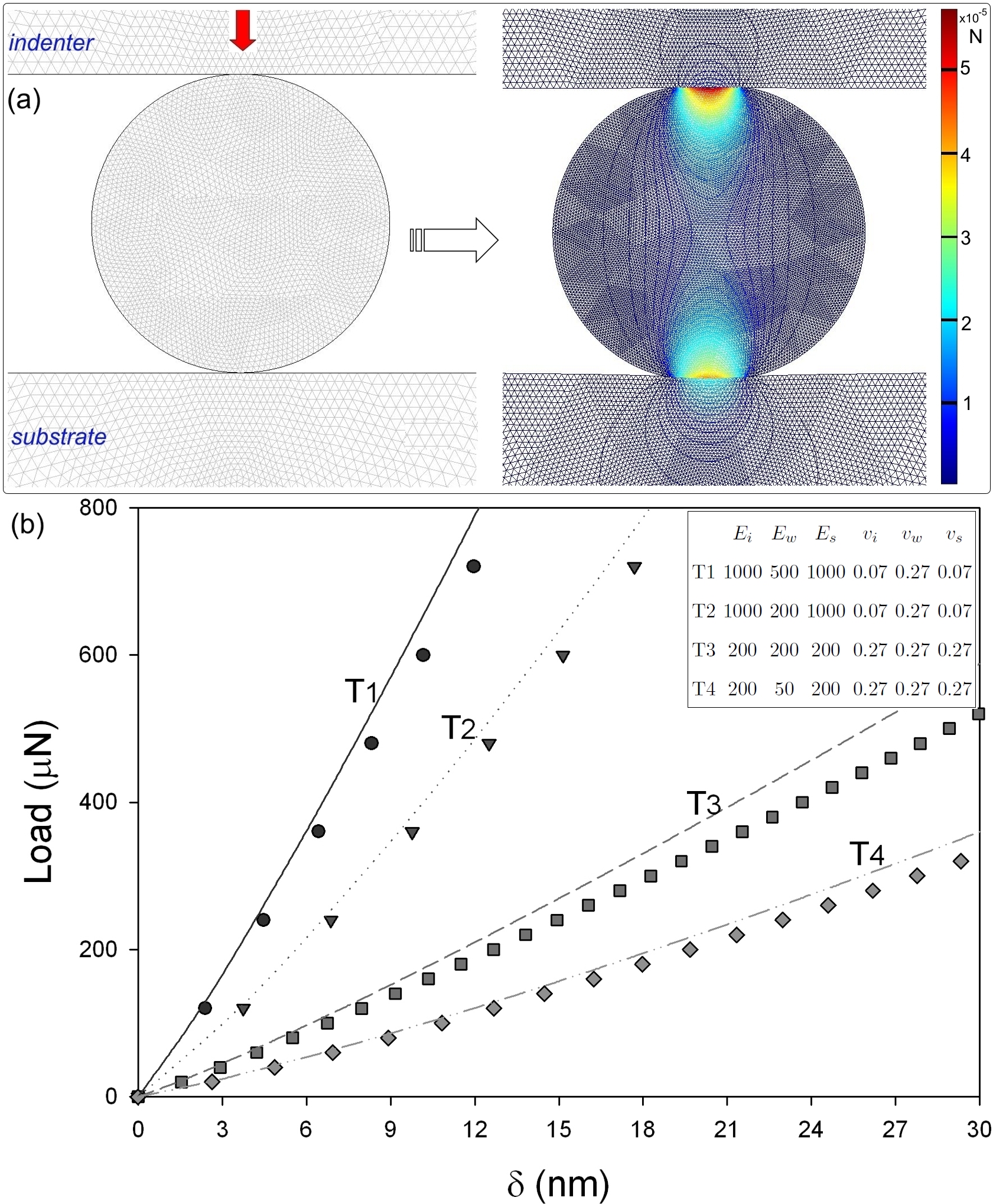}}
\caption{\label{fig:theovsimu}
(Color online) (a) \textit{Left:} Geometry meshes used in our FE indentation simulations. Color scale represents the distribution of strain energy (N$\cdot$m/m). A real-time video-recording of simulation is available online \cite{movie}. (b) Comparison between the load-displacement curve from Eq.\,\ref{eq:8} (lines) and that from simulations (symbols) with 4 different sets of parameters as that shown in the inset: Young's modulus $E$ (GPa) and Poisson's ratio $v$ ($h_{i}=h_{s}=100R=10$ $\mu$m).}
\end{figure}

Figure\,\ref{fig:theovsimu} shows a comparison between the compression experiment, the theoretical prediction and the finite-element (FE) simulation. It can be seen that the experimental elastic unloading curve roughly follows the track predicted by the simulation using the effective Young's modulus and Poisson's ratio estimated with EBSD. At the small deformation range, the load-displacement ratio predicted by the Hertzian model is close to that of the experiment and simulation. The difference becomes more important when the deformation increases.

The stability of this simplified Hertzian model has been checked by using FE simulations \cite{Comsol}. Our experimental configuration (flat punch + wire + substrate) was rebuilt in a plane-strain Lagrangian model applied to 4 different sets of material parameters (see Table 1). A vertical load was applied progressively to the top of the indenter (see Figure\,\ref{fig:schema} \textit{right}) and the deformation was then measured to compare with that predicted by Eq.\,\ref{eq:8}. The comparison of the $P-\delta$ curves is shown in Figure\,\ref{fig:theovsimu}. We can see that the stiffness increases with increasing elastic modulus of the materials. The stiffness obtained by simulations is sightly lower than that predicted analytically when the deformation is small. However, the difference becomes more significant in the case of large deformation. This may be due to the fact that the $R>>a$ Hertzian assumption is no longer valid in the large deformation case. 

In conclusion, in-situ SEM diametrical compression at the nano-scale has been introduced to measure the elastic modulus of individual nanowires with cylindrical cross-sections. The experimental diametral compression load-displacement response of a 250 nm diameter single crystal Si nanowire was compared to an analytical solution. The analytical solution of the three element contact mechanics problem was derived from Hertzian theory where the geometry of the experiment is simplified as the compression of an infinite long cylinder (nanowire) between two infinite half-spaces, the diamond flat punch and the silicon wafer respectively. Well characterized materials (diamond, silicon wafer and nanowire) were chosen for the three contact elements in order to have known material property values to feed into the analytical model. The comparison between the experimental measurement and the analytical solution showed good agreement and thus the feasibility of such experiment to determine the lateral elastic modulus of nanowires. The stability and reliability of the analytical solution was proven by comparing the load-displacement response obtained by the analytical model with the response obtained by FE numerical simulation applied to four different set of material parameters.

\end{document}